# A Simple Online Parameter Estimation Technique with Asymptotic Guarantees

Hien D. Nguyen

**Abstract**

In many modern settings, data are acquired iteratively over time, rather than all at once. Such settings are known as online, as opposed to offline or batch. We introduce a simple technique for online parameter estimation, which can operate in low memory settings, settings where data are correlated, and only requires a single inspection of the available data at each time period. We show that the estimators—constructed via the technique—are asymptotically normal under generous assumptions, and present a technique for the online computation of the covariance matrices for such estimators. A set of numerical studies demonstrates that our estimators can be as efficient as their offline counterparts, and that our technique generates estimates and confidence intervals that match their offline counterparts in various parameter estimation settings.

## I. Introduction

Let $\mathbf{x} = \{\boldsymbol{x}_i\}_{i=1}^N$ be a realization of a random sample $\mathbf{X} = \{\boldsymbol{X}_i\}_{i=1}^N$ of $N \in \mathbb{N}$ individual observations, where each $\boldsymbol{X}_i \in \mathbb{R}^d$ ($i \in [N]$; $[N] = \{1, \ldots, N\}$; $d \in \mathbb{N}$) arises from some marginal probability distribution $F$. Let $\boldsymbol{\theta} = \boldsymbol{\theta}(F) \in \mathbb{R}^p$ ($p \in \mathbb{N}$) be a parameter of $F$. The estimation of $\boldsymbol{\theta}$ from $\mathbf{x}$ is classically referred to as point estimation and is among the most important problems in statistics [1].

In the general setting, all of $\mathbf{x}$ is revealed simultaneously to the analyst in a batch fashion. This setting is known as offline learning. The alternative to offline learning is online learning, where the $N$ individual observations of $\mathbf{x}$ are revealed partially over some $T \in \mathbb{N}$ time periods, and where one is required to make an estimation of $\boldsymbol{\theta}$ at each time period $t \in [T]$.

The online learning context has been explored throughout the machine learning, signal processing, and statistics literature. Details of online machine learning of classifiers can be found in [2] and [3, Ch. 21]. A study of online learning of linear regression models can be found in [4, Sec. 8.7] and [5, Ch. 11]. A general treatment of online learning, concentrating on the passive-aggressive learning [6] and adaptive regularization of weight vectors methods [7], can be found in [8, Ch. 31]. Software implementations and further literature reviews of online algorithms can be found in [9] and [10].

In [9], it is noted that the online learning environment has three fundamental requirements that makes it different than that of the batch setting. Namely, (i) the observations are processed iteratively in time and are generally inspected only once,

Department of Mathematics and Statistics, La Trobe University, Bundoora, 3086 VIC, Australia (Email: h.nguyen5@latrobe.edu.au).

(ii) there is a limitation in space that inhibits the conversion of the online problem into an equivalent offline problem, (iii) estimation—given new observations—must be fast in comparison to the equivalent batch setting.

In the spirit of [9], we consider the following online learning setting. Firstly, we wish to estimate $\boldsymbol{\theta}$ from the $N$ observations of the sample $\mathbf{x}$, which are revealed over $T$ time periods in increments of $\nu \in \mathbb{N}$ (i.e. $N = \nu T$). Secondly, the processing unit that is required to make an estimation of $\boldsymbol{\theta}$ at each time point $t$ only has space for $\nu \le n < N$. Thirdly, the random variables $\boldsymbol{X}_i$ are allowed to arise from a stationary $m$-dependent process for some $m \in \mathbb{N}$ (cf. [11]) and there is enough space to store one estimate $\bar{\boldsymbol{\theta}}_{t-1} \in \mathbb{R}^p$, $M+1$ estimates $\hat{\boldsymbol{\theta}}_{t-j} \in \mathbb{R}^p$, and $M+1$ covariance matrices $\hat{\mathbf{V}}_{t-1}^{(-j)} \in \mathbb{R}^{p \times p}$ ($j \in [M] \cup \{0\}$) at each time period $t$, where $M \ge \max\{m/\nu, n/\nu - 1\}$. Lastly, there is sufficient space in the processing unit in order to perform all of the necessary computational operations on the $n$ available observations.

The approach that we utilize to solve the stated problem arises from our research into chunked-and-averaged (CA) estimators for non-IID (independent and identically distributed) data [12]. The CA estimators—for batch estimation in the IID setting—have been explored by [13] and [14] for regression problems, and [15] and [16] in the general univariate parameter estimation setting. Computation benefits of CA estimators in both the parallel and single processor settings have been considered in [17, Ch. 13].

The outcome of this letter is an online technique for computing sequences of estimates $\bar{\boldsymbol{\theta}}_t$ for any parameter $\boldsymbol{\theta}$, at each time period $t$, which are asymptotically normal with mean $\boldsymbol{\theta}$ and some covariance matrix $\bar{\mathbf{V}}_t$ that we can estimate via some covariance estimator $\hat{\bar{\mathbf{V}}}_t$. Maintaining the spirit of [9], we can compute our estimate and its estimated covariance matrix via algorithms that only inspects the sample of $n$ observations once, at each time $t$. Along with the presentation and theoretical justification of our online technique, we also conduct a set of numerical studies to demonstrate its application.

The letter proceeds as follows. The algorithm and its statistical properties are derived in Section II. A set of numerical studies are presented in order to demonstrate successful implementations of the algorithm in Section III. Conclusions are drawn in Section IV.

## II. MAIN RESULTS

Let $\mathbf{Y} = \{\boldsymbol{Y}_i\}_{i=-\infty}^{\infty}$ be a random sequence, where $\boldsymbol{Y}_i \in \mathbb{R}^q$ ($q \in \mathbb{N}$). Define $\mathcal{B}_n^{n+k}$ be to the Borel $\sigma$-field that is generated by the set $\{\boldsymbol{Y}_n, \boldsymbol{Y}_{n+1}, \ldots, \boldsymbol{Y}_{n+k}\}$. Further, define

$$\alpha(\mathcal{A}, \mathcal{B}) = \sup_{\{A \in \mathcal{A}, B \in \mathcal{B}\}} |\mathbb{P}(A \cap B) - \mathbb{P}(A)\mathbb{P}(B)|,$$

where $\mathcal{A}$ and $\mathcal{B}$ are Borel $\sigma$-fields, and let $\alpha(k) = \sup_n \alpha\left(\mathcal{B}_{-\infty}^n, \mathcal{B}_{n+k}^\infty\right)$. We say that $\mathbf{Y}$ is $\alpha$-mixing (or strong mixing) with rate $-s$ if $\lim_{k \to \infty} \alpha(k) \to 0$ and $\alpha(k) = O(k^{-s-\epsilon})$, for some $\epsilon > 0$ [18, Sec. 3.4].

Define $\mathbf{Y}$ to be an $M$-dependent sequence if for each $i$, $\boldsymbol{Y}_i$ is independent of $\boldsymbol{Y}_j$ if and only if $|i - j| > M \ge 0$. For an $M$-dependent sequence, we observe that $\alpha(k) = 0$ for $k > M$ by definition. Furthermore, for any $\epsilon > 0$ and $s > 0$, we also observe that $\alpha(k) = 0 < k^{-s-\epsilon}$, for $k > M$. Thus, an $M$-dependent sequence can be concluded to be strong mixing with a rate $-s \to -\infty$.

Recall that the processing unit used for computation can hold $n$ sample observations, which are obtained at a rate of $\nu$ at each time period. For sufficiently large $t$ (to fill up the $n$ available spaces for observations), let $\mathbf{x}_t = \{\boldsymbol{x}_{t1}, \ldots, \boldsymbol{x}_{tn}\}$ be the realized sample of $n$ observations. Since $\nu$ new observations are acquired at time $t+1$, we must discard the $\nu$ oldest observations from $\mathbf{x}_t$. Let $\boldsymbol{y}_{t+1,1}, \ldots, \boldsymbol{y}_{t+1,\nu}$ be the $\nu$ new observations that are obtained. Thus, at period $t+1$, $\mathbf{x}_{t+1} = \{\boldsymbol{x}_{t,\nu+1}, \ldots, \boldsymbol{x}_{tn}, \boldsymbol{y}_{t+1,1}, \ldots, \boldsymbol{y}_{t+1,\nu}\}$. Let $\mathbf{X}_t$ be the random version of $\mathbf{x}_t$. We have the following result regarding the sequence $\{\mathbf{X}_t\}_{t=-\infty}^{\infty}$.

**Proposition 1.** *If $\mathbf{X} = \{\boldsymbol{X}_i\}_{i=-\infty}^{\infty}$, where each $\boldsymbol{X}_i$ is $m$-dependent in the index $i$, and*

$$\mathbf{X}_t = \{\boldsymbol{X}_{t1}, \ldots, \boldsymbol{X}_{tn}\} = \{\boldsymbol{X}_{i+1}, \ldots, \boldsymbol{X}_{i+n}\},$$

*for some $i$, is a consecutive subset of $n$ observations from $\mathbf{X}$, whereby $\mathbf{X}_{t+1}$ is obtained from $\mathbf{X}_t$ via the iterations*

$$\mathbf{X}_{t+1} = \{\boldsymbol{X}_{t,\nu+1}, \ldots, \boldsymbol{X}_{tn}, \boldsymbol{X}_{i+n+1}, \ldots, \boldsymbol{X}_{i+n+\nu}\}, \tag{1}$$

*for $1 \leq \nu \leq n$, then the sequence $\{\mathbf{X}_t\}_{t=-\infty}^{\infty}$ is $M$-dependent for any $M \geq \max\{m/\nu, n/\nu - 1\}$.*

*Proof:* The dependence between each $\mathbf{X}_t$ and $\mathbf{X}_u$, for $t, u \in \mathbb{Z}$ can occur via two mechanisms, that is, the $m$-dependence from $\mathbf{X}$ or the dependence from the overlap of the iterations (1). At each time point $t$, the leftmost $\nu$ indices (in $i$) are discarded, and since a discarded observation can only be dependent on the $m$ observations to its right, by the pigeonhole principle, a discarded observation from $\mathbf{X}_t$ can only be dependent on $\mathbf{X}_u$ for $u \leq m/\nu$. Via the pigeonhole principle, you can fit at most $\lceil n/\nu \rceil$ ($\lceil \cdot \rceil$ is the ceiling operator) sets of $\nu$ observations in a space of $n$, it only requires $\lceil n/\nu \rceil - 1$ iterations of process (1) in order for two samples to share only the rightmost $\nu$ observations (with respect to index $i$) in common. Thus $\mathbf{X}_t$ and $\mathbf{X}_{\lceil n/\nu \rceil}$ are independent, which implies that $\{\mathbf{X}_t\}_{t=-\infty}^{\infty}$ is also $(n/\nu - 1)$-dependent. Since $\{\mathbf{X}_t\}_{t=-\infty}^{\infty}$ must both be $(m/\nu)$- and $(n/\nu - 1)$-dependent, we obtain the result via the definition of $M$-dependence. ∎

Let $\hat{\boldsymbol{\Theta}}_t = \hat{\boldsymbol{\theta}}(\mathbf{X}_t)$ be the estimator of $\boldsymbol{\theta}$ at time period $t$, and let $\hat{\boldsymbol{\theta}}_t = \hat{\boldsymbol{\theta}}(\mathbf{x}_t)$ be its realization. By the continuous mapping result of [18, Thm. 3.49], the sequence $\left\{\hat{\boldsymbol{\Theta}}_t\right\}_{t=-\infty}^{\infty}$ is $M$-dependent for any $M \geq \max\{m/\nu, n/\nu - 1\}$, given that it is a measurable function mapping into $\mathbb{R}^p$.

Define $\boldsymbol{\mu}_t = \mathbb{E}\left(\hat{\boldsymbol{\Theta}}_t\right)$ and $\mathbf{V}_t = \text{Cov}\left(\hat{\boldsymbol{\Theta}}_t\right)$, respectively. Further, let $\boldsymbol{\mu}_t^\top = (\mu_{t1}, \ldots, \mu_{tp})$ and $\hat{\boldsymbol{\Theta}}_t^\top = \left(\hat{\Theta}_{t1}, \ldots, \hat{\Theta}_{tp}\right)$, and define $\bar{\boldsymbol{\Theta}}_t = t^{-1} \sum_{u=1}^{t} \hat{\boldsymbol{\Theta}}_u$ to be the CA estimator of $\boldsymbol{\theta}$ at time $t$. Here $\top$ is the transposition operator. The following result can be interpreted from [12, Prop. 18].

**Theorem 2.** *Assume that the sequence $\left\{\hat{\boldsymbol{\Theta}}_t\right\}_{t=1}^{T}$ is $\alpha$-mixing with size $-s$, where $\mathbb{E}\left|\hat{\Theta}_{tj} - \mu_{tj}\right|^{2s/(s-1)} < \Delta < \infty$ for some $s > 0$, uniformly for all $t \in [T]$ and $j \in [p]$. If $\bar{\boldsymbol{\mu}}_t = t^{-1} \sum_{u=1}^{t} \boldsymbol{\mu}_u$ and $\bar{\mathbf{V}}_t = t^{-1} \sum_{u=1}^{t} \mathbf{V}_u$, and if $\bar{\mathbf{V}}_t$ is strictly positive-definite for all sufficiently large $t$, then $\bar{\boldsymbol{\Theta}}_t$ is asymptotically normal with mean vector $\bar{\boldsymbol{\mu}}_t$ and covariance matrix $\bar{\mathbf{V}}_t$.*

Since $\left\{\hat{\boldsymbol{\Theta}}_t\right\}_{t=1}^{T}$ is $M$-dependent, we can take $-s \to \infty$ which implies that it is sufficient to assume the existence of second moments in order to validate the hypothesis of Theorem 2. Note that the asymptotic normality result implies that $t^{1/2} \bar{\mathbf{V}}_t^{-1/2} \left(\bar{\boldsymbol{\Theta}}_t - \bar{\boldsymbol{\mu}}_t\right)$ converges in law to a standard multivariate normal random variable. Further notice, that the pivot of the asymptotic normality is about $\bar{\boldsymbol{\mu}}_t$ and not $\boldsymbol{\theta}$.

Consider the expansion:

$$t^{1/2}\bar{\mathbf{V}}_t^{-1/2}\left(\bar{\boldsymbol{\Theta}}_t - \boldsymbol{\theta}\right) \qquad (2)$$
$$= t^{1/2}\bar{\mathbf{V}}_t^{-1/2}\left(\bar{\boldsymbol{\Theta}}_t - \bar{\boldsymbol{\mu}}_t\right) + t^{1/2}\bar{\mathbf{V}}_t^{-1/2}\left(\bar{\boldsymbol{\mu}}_t - \boldsymbol{\theta}\right).$$

From (2), we have the result that we seek if the final term goes to zero. That is, if $\bar{\boldsymbol{\mu}}_t - \boldsymbol{\theta} = o\left(\bar{\mathbf{V}}_t^{1/2} t^{-1/2}\right)$ (A), then $\bar{\boldsymbol{\Theta}}_t$ is asymptotically normal with mean vector $\boldsymbol{\theta}$ and covariance matrix $\bar{\mathbf{V}}_t$. A condition that implies this is for each $\hat{\boldsymbol{\Theta}}_t$ to be an unbiased estimator of $\boldsymbol{\theta}$.

We shall work with the assumption that (A) holds for remainder of the letter. Given this assumption, $\bar{\boldsymbol{\Theta}}_t$ is a consistent estimator of $\boldsymbol{\theta}$ and a simple scheme for the online computation of the estimate $\bar{\boldsymbol{\theta}}_t = t^{-1}\sum_{u=1}^t \hat{\boldsymbol{\theta}}_u$, given only the newly computed estimate $\hat{\boldsymbol{\theta}}_t = \boldsymbol{\theta}(\mathbf{x}_t)$ and the previous CA estimate $\bar{\boldsymbol{\theta}}_{t-1}$, is via the update rule:

$$\bar{\boldsymbol{\theta}}_t = \frac{(t-1)\bar{\boldsymbol{\theta}}_{t-1} + \hat{\boldsymbol{\theta}}_t}{t}. \qquad (3)$$

Unfortunately, unlike (3), a simple expression and update scheme for the covariance matrix $\bar{\mathbf{V}}_t$ is not available under only the assumption of (A). Since $\{\hat{\boldsymbol{\Theta}}_t\}_{t=1}^T$ is $M$-dependent, we can make the structural assumption that $\mathbb{E}\left(\hat{\boldsymbol{\Theta}}_t - \boldsymbol{\mu}_t | \mathcal{B}_{-\infty}^{t-M-1}\right) = \mathbf{0}$ (cf. [18, Sec. 6.3]). A further assumption of unbiasedness implies the expansion:

$$\bar{\mathbf{V}}_t = t^{-1}\sum_{u=1}^t \mathbb{E}\left[\left(\hat{\boldsymbol{\Theta}}_u - \boldsymbol{\theta}\right)\left(\hat{\boldsymbol{\Theta}}_u - \boldsymbol{\theta}\right)^\top\right] \qquad (4)$$
$$+ t^{-1}\sum_{j=1}^M \sum_{u=j+1}^t \mathbb{E}\left[\left(\hat{\boldsymbol{\Theta}}_u - \boldsymbol{\theta}\right)\left(\hat{\boldsymbol{\Theta}}_{u-j} - \boldsymbol{\theta}\right)^\top\right]$$
$$+ t^{-1}\sum_{j=1}^M \sum_{u=j+1}^t \mathbb{E}\left[\left(\hat{\boldsymbol{\Theta}}_{u-j} - \boldsymbol{\theta}\right)\left(\hat{\boldsymbol{\Theta}}_u - \boldsymbol{\theta}\right)^\top\right].$$

Matrix (4) can in turn be estimated by

$$\hat{\bar{\mathbf{V}}}_t = \hat{\bar{\mathbf{V}}}_t^{(0)} + \sum_{j=1}^M \hat{\bar{\mathbf{V}}}_t^{(-j)} + \sum_{k=1}^M \hat{\bar{\mathbf{V}}}_t^{(-j)\top} \qquad (5)$$

where $\hat{\bar{\mathbf{V}}}_t^{(-j)} = \sum_{u=j+1}^t \left(\hat{\boldsymbol{\theta}}_u - \bar{\boldsymbol{\theta}}_t\right)\left(\hat{\boldsymbol{\theta}}_{u-j} - \bar{\boldsymbol{\theta}}_t\right)^\top$, for $j \in [M] \cup \{0\}$.

We can adapt the recursive formula of [19] to obtain the following online update rule for computing $\hat{\bar{\mathbf{V}}}_t^{(-j)}$ in (5):

$$\mathbf{V}_t^{(-j)} = \frac{t-1}{t}\left[\mathbf{V}_{t-1}^{(-j)} + t^{-1}\left(\hat{\boldsymbol{\theta}}_t - \bar{\boldsymbol{\theta}}_{t-1}\right)\left(\hat{\boldsymbol{\theta}}_{t-j} - \bar{\boldsymbol{\theta}}_{t-1}\right)^\top\right]. \qquad (6)$$

Finally, under the assumption that $\hat{\boldsymbol{\Theta}}_t$ is unbiased at each time period $t$, the matrix (5)—evaluated at the random observations, instead of the realizations—is a consistent estimator of (4) via a Slutsky-type argument and a law of large numbers for stationary ergodic sequence (e.g. [18, Thm. 3.34]).

III. NUMERICAL STUDIES

We perform a set of three numerical studies. In all three studies, both $\nu = 100$ and $T = 100$, implying that $N = \nu T = 10000$. In the first study (S1), we simulate bivariate IID data $\mathbf{X}_i^\top = (X_i, Y_i)$, where $X_i$ is drawn from the standard normal distribution, and $Y_i = \beta_1 + \beta_2 X_i + E_i$, $\beta_1 = 0$, $\beta_2 = 2$ and $E_i$ is standard normal. Here, we set $n = 500$, and thus $\{\mathbf{X}_t\}_{t=1}^T$ can be considered $M$-dependent, where $M = 4$. In S1, we wish to estimate $\boldsymbol{\theta}^\top = (\beta_1, \beta_2)$, which can be achieved via ordinary least squares (OLS; see [4, Sec. 8.4]). Let $\hat{\boldsymbol{\Theta}}_t$ be the OLS estimator computed on sample $\mathbf{X}_t$. Via the usual, Gauss-Markov argument, it is known that the OLS estimator is unbiased for $\boldsymbol{\theta}$, thus our assumptions for the use of (3) and (6) are validated, when we construct our CA estimator using OLS estimators.

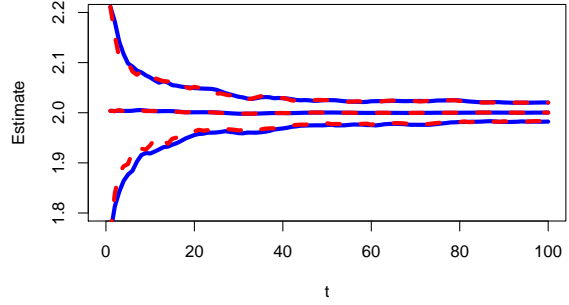

Figure 1. Mean and PIs of estimates for $\beta_2$ in S1. Solid and dotted lines correspond to the CA and batch estimates.

In the second study (S2), we simulate univariate IID data $X_i$ from a Laplace distribution with median $\lambda_1 = -1$ and scale parameter $\lambda_2 = 1$, which implies that the variance is $2\lambda_2^2 = 2$. Here we set $n = 200$, which implies that $\{\mathbf{X}_t\}_{t=1}^T$ is $M$-dependent, where $M = 1$. In S2, we wish to estimate $\boldsymbol{\theta}^\top = (\lambda_1, \lambda_2^2)$, which can be achieved via the sample median and the variance divided by two. Let $\hat{\boldsymbol{\Theta}}_t^\top = (\mathrm{med}_t, S_t^2/2)$, where $\mathrm{med}_t$ is the sample median and $S_t^2$ is the sample variance over the set $\mathbf{X}_t$, respectively. It is well-known that the median is an unbiased estimator of the centre of any symmetric distribution and that the sample variance is an unbiased estimator for the population variance, under IID sampling. Thus the assumptions for (3) and (6) are again validated, when we construct our CA estimator using the aforementioned statistics.

In the third study (S3), we simulate univariate data $X_i$ from a first-order moving average process of the form $X_i = \rho_1 + E_i + \rho_2 E_{i-1}$, where for each $i$, $E_i$ is generated from a standard normal distribution. Here, $\rho_1 = 0$ and $\rho_2 = 0.5$. In this study, we set $n = 100$, which implies that $\{\mathbf{X}_t\}_{t=1}^T$ is 1-dependent, since $\mathbf{X}$ is 1-dependent. In S3, we wish to estimate $\boldsymbol{\theta}^\top = (\rho_1, \rho_2)$, which we can do via the maximum likelihood estimator (MLE). Unfortunately, the MLE $\hat{\boldsymbol{\Theta}}_t$ is only asymptotically unbiased in this setting, thus our assumptions are not entirely validated here; see [20, Ch. 8].

In order to make an assessment regarding the success of the online CA estimator (3), we repeat each study 100 times and obtain the mean and the 95% percentile interval (PI) of the estimates $\hat{\boldsymbol{\theta}}_t$, for each $t \in [T]$. We then compare these means and PIs to that of the equivalent batch estimates obtained using all of the $\nu t$ available observations, at each time period $t$. For example, in S1, we would compare the CA estimates against the OLS estimates using all of the available observations $\nu t$ at each time period $t$. Figures 1–3 display the means and PIs for the estimates of $\beta_2$, $\lambda_1$, and $\rho_2$, respectively.

From Figures 1–3, we observe that the CA estimates have similar mean performances as well as variation when compared to their batch counterparts. This implies that online estimation via (3) is as efficient as offline estimation, using the same base estimators. Indeed it has been proved that in the IID case and when $\{\mathbf{X}_t\}_{t=1}^T$ are not overlapping (i.e. $\nu = n$), that the CA estimator is efficient under mild assumptions (cf. [15], [12]). Here however, it is more surprising, since correlated data generally result in more dispersed estimates. In particular, it is surprising that such efficiency is matched in S1, since the OLS is known to be the minimum variance unbiased estimator.

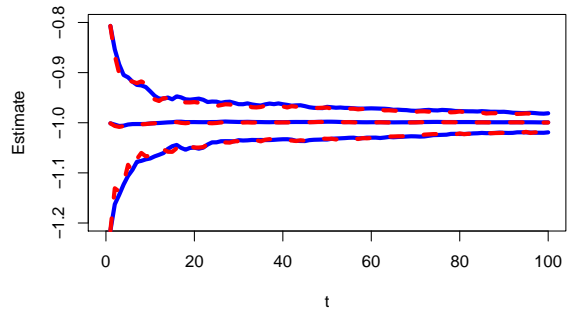

Figure 2. Mean and PIs of estimates for $\lambda_1$ in S2. Solid and dotted lines correspond to the CA and batch estimates.

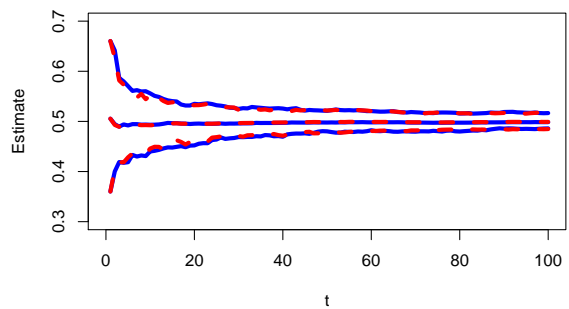

Figure 3. Mean and PIs of estimates for $\rho_2$ in S3. Solid and dotted lines correspond to the CA and batch estimates.

Asymptotic normality results for the OLS estimator, sample median, and MLE for moving average models are well-known. Using these results, we construct 95% asymptotic confidence intervals (CIs) for single instances of the batch estimates, across all $t$, and compare them to the asymptotic CIs that are constructed via (3) and (5) for the respective estimates of $\beta_2$, $\lambda_1$, and $\rho_2$. The CIs are displayed in Figures 4–6.

From Figures 4–6, we observe that (3) and (5) can be used to compute estimates and CIs that match those that can be constructed via offline estimation, especially for higher values of $t$. This is a good result, especially in the case of S3, where the assumptions for the use of (5) are not entirely met.

We finally note that in each of the studies, we can also conduct a Shapiro-Wilk test for normality ([21], [22]) on each sample of 100 replicates of estimates for $\beta_1$, $\beta_2$, $\lambda_1$, $\lambda_2^2$, $\rho_1$, and $\rho_2$, at time period $t = T$. The respective $p$-values are 0.4138, 0.7813, 0.8978, 0.7443, 0.4788, and 0.3254, which indicates that the asymptotia of Theorem 2 appears reached in each of the studies.

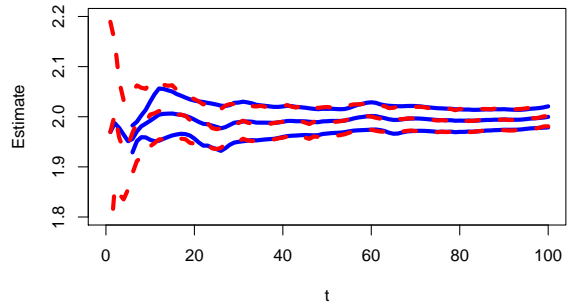

Figure 4. CIs for estimates of $\beta_2$ in S1. Solid and dotted lines correspond to the CA and batch estimates.

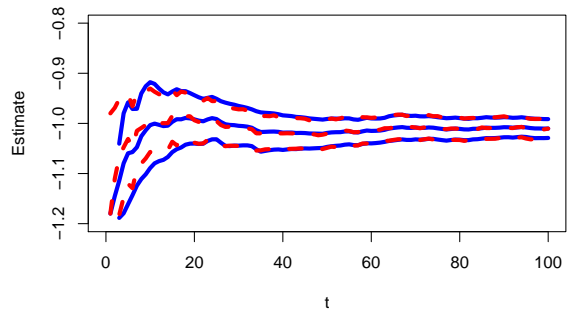

Figure 5. CIs for estimates of $\lambda_1$ in S2. Solid and dotted lines correspond to the CA and batch estimates.

## IV. Conclusions

We have presented a simple online estimation technique that is based on the CA estimators and can be applied in general settings where parameter estimation is required. The constructed online estimators can be shown to be asymptotically normal under generous assumptions and their computation, and the computation of their covariance matrices can be conducted by only making one inspection of the available data, at each time point. Furthermore, via some numerical studies, we establish that our online technique can produce estimators that are as efficient as their offline counterparts, and that can produce estimates and confidence intervals that match those of their batch versions, even when the strict assumptions of our theoretical results are not met.

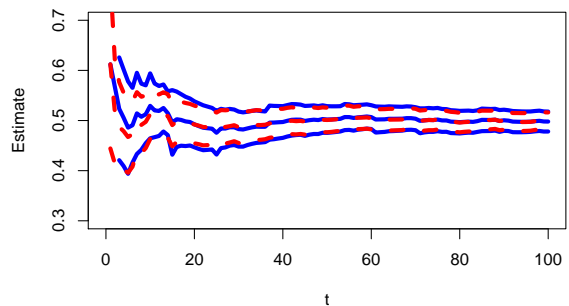

Figure 6. CIs for estimates of $\rho_2$ in S3. Solid and dotted lines correspond to the CA and batch estimates.